\documentclass[conference]{IEEEtran}
\IEEEoverridecommandlockouts
\usepackage{cite}
\usepackage{amsmath,amssymb,amsfonts}
\usepackage{algorithmic}
\usepackage{graphicx}
\usepackage{textcomp}
\usepackage{xcolor}
\usepackage{orcidlink}
\usepackage{multicol, multirow}

\def\BibTeX{{\rm B\kern-.05em{\sc i\kern-.025em b}\kern-.08em
    T\kern-.1667em\lower.7ex\hbox{E}\kern-.125emX}}
\begin{document}

\title{\textbf{POLARON}: Precision-aware On-device Learning and Adaptive Runtime-cONfigurable AI acceleration\\ 
\thanks{This work was supported by the Special Manpower Development Program for Chip to Startup (SMDP-C2S), Ministry of Electronics and Information Technology (MeitY), Govt. Of India. We acknowledge the use of ChatGPT (OpenAI, 2025) for proofreading, grammar correction, and language refinement to improve readability in the early drafts of this manuscript. All technical content, analyses, and results are solely the original contributions of the authors.

}
}

\author{
    \IEEEauthorblockN{Mukul Lokhande\IEEEauthorrefmark{1}\orcidlink{0009-0001-8903-5159}, 
    Santosh Kumar Vishvakarma\IEEEauthorrefmark{1}\orcidlink{0000-0003-4223-0077}}
    \IEEEauthorblockA{\IEEEauthorrefmark{1}NSDCS Research Group, Indian Institute of Technology Indore, India}
    Email: skvishvakarma@iiti.ac.in \textbf{(Corresponding Author)}
}

\maketitle

\begin{abstract}
The increasing complexity of AI models requires flexible hardware capable of supporting diverse precision formats, particularly for energy-constrained edge platforms. This work presents PARV-CE, a SIMD-enabled, multi-precision MAC engine that performs efficient multiply-accumulate operations using a unified data-path for 4/8/16-bit fixed-point, floating-point, and posit formats. The architecture incorporates a layer-adaptive precision strategy to align computational accuracy with workload sensitivity, optimizing both performance and energy usage. PARV-CE integrates quantization-aware execution with a reconfigurable SIMD pipeline, enabling high-throughput processing with minimal overhead through hardware-software co-design. The results demonstrate up to 2× improvement in PDP and 3× reduction in resource usage compared to SoTA designs, while retaining accuracy within 1.8\% FP32 baseline. The architecture supports both on-device training and inference across a range of workloads, including DNNs, RNNs, RL, and Transformer models. The empirical analysis establish PARV-CE incorporated POLARON as a scalable and energy-efficient solution for precision-adaptive AI acceleration at edge.
\end{abstract}


\section{Introduction}

\begin{figure}[!t]
  \centering
  \includegraphics[width=0.85\columnwidth]{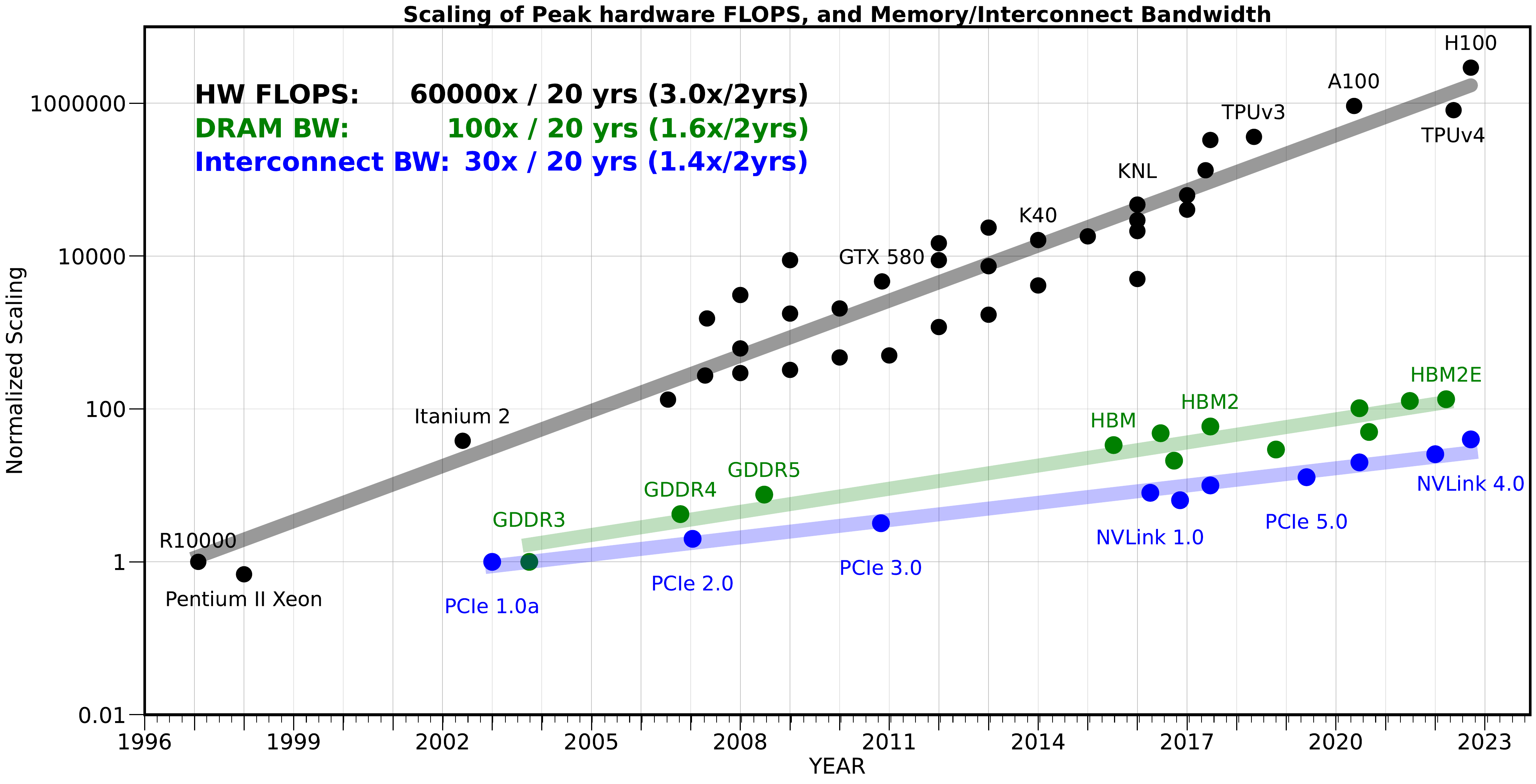}
  \caption{Decades-long growth in hardware FLOPS, DRAM, and interconnect bandwidth, emphasizing the breakdown of Dennard Scaling and the need for precision-aware compute under the Roofline model. Adapted from \cite{AI-Memory_wall}}
  \label{fig:hw-scaling}
  \vspace{-7mm}
\end{figure}

\IEEEPARstart{T}{he} Artificial Intelligence (AI) landscape has witnessed rapid growth, fueled by the emergence of generative applications such as image manipulation, reinforcement learning (RL), and text-to-image/video generation\cite{AI-accl1, AI-accl2}. Initially designed to automate fundamental tasks such as image classification and data augmentation, particularly deep neural networks (DNNs) has evolved into a powerful tool that augments human creativity and enables intelligent decision making. This progress has accelerated the adoption of AI on mobile and Internet of Things (IoT) platforms, where stringent power, memory, and bandwidth constraints\cite{GPU-opt-nvidia,AI-Memory_wall} pose significant challenges for real-time inference and on-device learning as visible from Fig. \ref{fig:hw-scaling}. To meet these demands, Modern AI workloads increasingly rely on specialized accelerators offering improved energy efficiency, lower latency, and compact area footprints compared to traditional CPUs. Architectures such as TPUs, GPUs, and custom ASICs or FPGAs are co-optimized across hardware and software layers using model compression, architectural innovations, and alternate number systems\cite{Mega-mini}. Nonetheless, deploying within tight resource budgets of embedded platforms remains non-trivial, bottlenecked by transmission bandwidth.

\begin{table}[!b]
\vspace{-5mm}
\caption{State-of-the-art (SoTA) computing approaches and respective features in multiply-accumulate compute engines}
\label{sota-comp}
\renewcommand{\arraystretch}{1.25}
\resizebox{\columnwidth}{!}{%
\begin{tabular}{c|c|c|c|c|c}
\hline
\textbf{Design} & \textbf{Computing} & \textbf{Bit-width} & \textbf{Datatypes} & \textbf{Limitations} & \textbf{Applications} \\ \hline
TVLSI'25\cite{Flex-PE} & Multi-Precision & 4/8/16/32 & FxP & Memory-bound & FPGA \\\hline
TCAD'25\cite{MACC-AMD} & Mixed-Precision & 4/8/16/32 & FP/BF/TF & Memory-bound & FPGA\\ \hline
TCAS-II'24\cite{RPE-TCASII'24} & Mixed-Precision & 2/4/8/16/32 & INT/FP/BF/TF & Memory-bound & HPC \\ \hline
HCS'24 (NVIDIA) & Mixed-Precision & 4/6/8/16 & FP/BF & - & GPU \\ \hline
TCAD'24\cite{Tan-TCAD'24} & - & 16/32 & FP/HFP/BF/TF & Compute-bound & HPC \\ \hline
TCAS-II'24\cite{Tan-TCASII'24} & Mixed-Precision & 16/32/64 & FP/TF/BF & Compute-bound & HPC \\ \hline
Micro'24 (AMD) & Mixed-Precision & 4/8/16/32 & INT/BF & Memory-bound & PC / Desktop \\ \hline
IoTJ'24\cite{RAMAN} & Mixed-Precision & 2/4/8 & FxP & Compute-bound & edge AI \\ \hline
JSSC'23\cite{Samsung-JSSC} & Multi-Precision & 4/8/16 & INT/FP & Compute-bound & Mobile AI \\ \hline
ESSCIRC'23 & Mixed-Precision & 4/8 & FP & Compute-bound & DNN Training \\ \hline
TCAS-II'22\cite{Posit-FP-VMAC-TCASII'22} & Multi-Precision & 8/16/32 & FP/Posit & Memory-bound & - \\ \hline
ISCA'21 (IBM) & Mixed-Precision & 2/4/8/16 & FP/FxP & - & DNNs \\ \hline
TC'20 & Multi-Precision & 1/2/4/8/16/32 & FP/FxP & Memory-bound & DNN Training \\ \hline
Proposed & Trans-Precision & 4/8/16 & FP/Posit/FxP & Runtime-adaptable & \begin{tabular}[c]{@{}c@{}}On-device Learning\\ AI workloads\end{tabular} \\ \hline
\end{tabular}}
\end{table}

Multiply-accumulate (MAC) operations dominate the computational workload in DNNs\cite{Flex-PE}, accounting for over 90\% of total operations particularly within Convolutional layers or multi-layer perceptron (MLP). For instance, the VGG-16 architecture requires approximately 15.5G MACs, whereas large transformer models such as GPT-3 demand over 6.7T MACs. Thus, the resource-efficient design and optimization of MAC units becomes critical to enhance system-level throughput. However, implementing multiplexed separate-precision data-paths incurs significant hardware overhead and leads to dark silicon\cite{Tan-TCASII'24, Tan-TVLSI'23}. Trans-Precision computing, with dynamic adjustment of precision formats and data widths, has proven effective in enhancing resource efficiency without compromising application accuracy, particularly by enabling selective use of lower-precision computations\cite{SIMD-mac-patent-intel, SIMD-MAC-USP-NVIDIA,ARM-MAC-USP}. Our work investigates hardware support for the same, to enable efficient on-device learning (ODL) acceleration under resource constraints across diverse AI workloads. Table~\ref{sota-comp} presents a comprehensive summary of recent commercial computing devices, with detailed analysis of integrated MAC units such as computing type, bit-width, data format, performance limitations, and target applications. Notably, the proposed work stands out as the first to offer runtime-adaptable trans-precision computation across all fixed-point (FxP), floating-point (FP), Posit formats and 4/8/16-bit precision. This versatility enables efficient deployment across a wide range of AI workloads, including DNNs, transformers, RL, and generative AI applications.

The primary contributions are summarized as follows:
\begin{itemize}

    \item \textbf{Precision-Aware Runtime-adaptive Vector Compute Element (PARV-CE)}: This work introduces enhanced-performance, precision-aware on-device learning CE, which supports adaptive, runtime-configurable throughput up to 16× for Var-FxP4 and Var-FP8 (E5M2, E4M3), 4× for Var-FxP8, Posit8, and BF16, and 1× for Var-FxP16, Posit16, and Var-FP16 (E5M10, E6M9). This is enabled through a resource-shared MAC architecture that maintains near 100\% hardware utilization.

    \item \textbf{WILD-QLite Quantization Algorithm:} A novel distribution-aware quantization framework that dynamically adjusts bit-widths based on workload sensitivity and layer criticality to maximize hardware efficiency. Co-designed with a configurable multi-port memory architecture, the algorithm enables parallel data access and optimized resource utilization, supporting low-precision training and inference with minimal accuracy degradation.

    \item \textbf{Resource-Efficient POLARON Edge-AI Engine Analysis:} An empirical performance evaluation of the enhanced POLARON Edge-AI engine is presented, highlighting the impact of runtime-adaptable precision formats and data widths on application performance and hardware resource utilization.
    
\end{itemize}


\section{Conceptualization : POLARON edge AI Engine}

\subsection{PARV-CE : MAC Compute Unit}
Several commercial efforts\cite{SIMD-mac-patent-intel, SIMD-MAC-USP-NVIDIA, ARM-MAC-USP} have focused on the development of SIMD-MAC units, supporting multi-precision product accumulation, with shared data-paths to enhance resource efficiency and throughput, for AI-ML applications. The prior works\cite{MACC-AMD, RPE-TCASII'24, Tan-TCAD'24, Tan-TCASII'24, Posit-FP-VMAC-TCASII'22} primarily decomposed compute data-path into multiple pipeline stages such as input processing, sign handling, exponent/regime processing, mantissa multiplication, output accumulation, and final output formatting depending on the precision involved, latency to be achieved and hardware efficiency. We designed our compute unit as a five-stage pipelined architecture, with similar processing phases - ranging from input pre-processing to final output handling, while focused on low-latency and area-efficiency. What differentiates this work is comprehensive support for diverse precision formats is Var-FxP4/8 for DNNs, RNNs, and reinforcement learning (RL); FP8 for large language models (LLMs); Posit-8 and Posit-16 for training; and BF16 for general-purpose AI computations enabling efficient and flexible on-chip execution across a wide range of AI workloads.

\begin{figure}[!t]
    \centering
    \includegraphics[width=0.8\columnwidth,height=68mm]{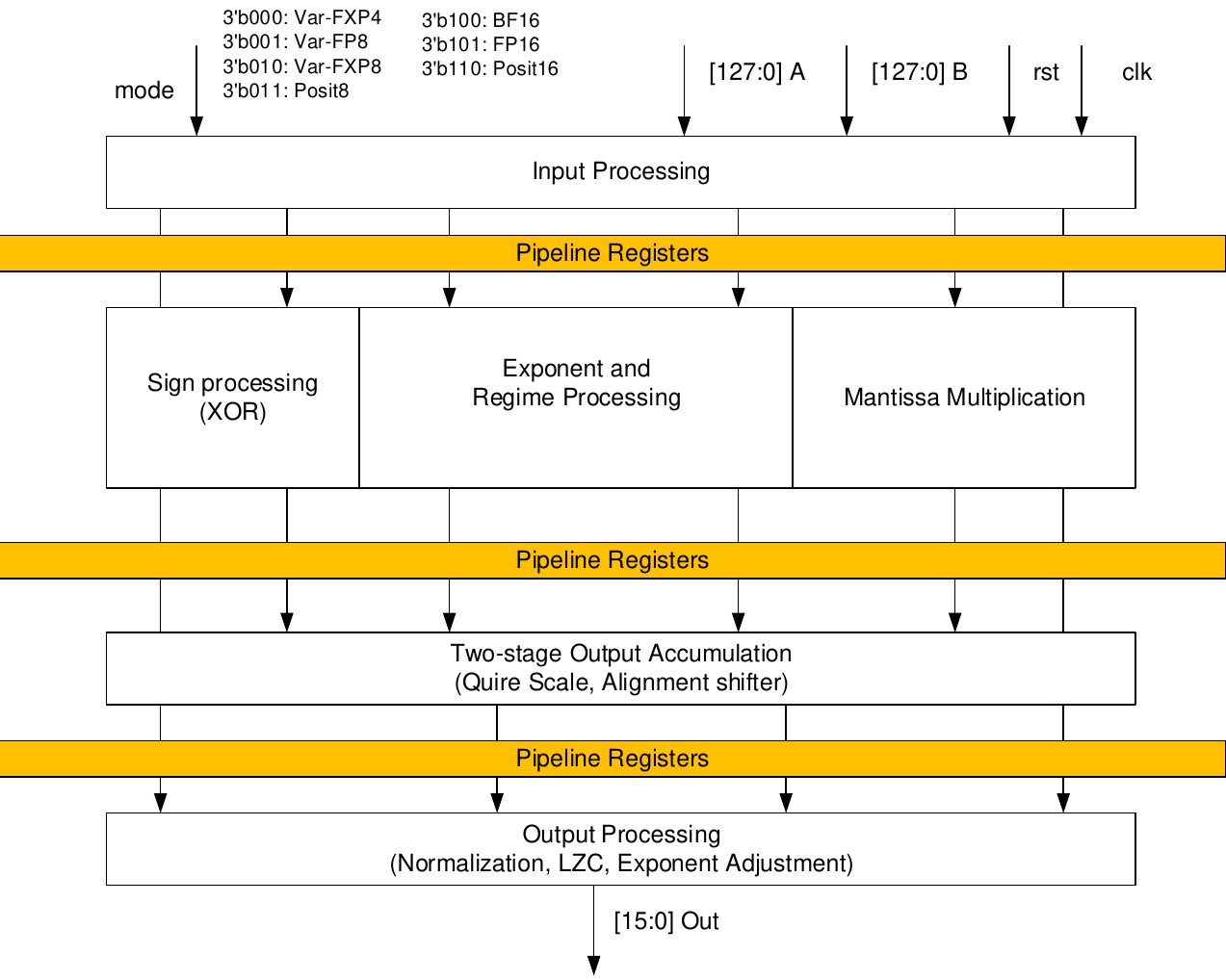}
    \vspace{-3mm}
    \caption{The detailed micro-architecture of 5-stage Precision-Aware Runtime-adaptive Vector Compute Element (PARV-CE) for enhanced-performance ON-device learning}
    \label{Fig-polaron}
    \vspace{-7 mm}
\end{figure}

\textbf{Variable Precision Processing:}
The Posit-16 fused MAC data-path is architected to support morphable SIMD-precision, enabling runtime adaptation to various input formats as per the specified operating modes. The design efficiently manages scale factors and minimizes hardware overhead, similar to \cite{Posit-FP-VMAC-TCASII'22}. The input processing block (Stage-I) performs the unpacking of operands by extracting the sign, exponent/regime, and mantissa fields, and route to respective computational units. To support variable bit-widths across different precision modes, a non-unified memory access scheme is employed using a configurable multi-port memory architecture. The architectural integration through an adaptive, precision-aware data pre-fetcher and configuration subset controller, ensuring efficient data handling and processing across modes.

\textbf{Runtime adaptive SIMD Multiplication:}
Stage II performs vectorized multiplication using 16x 4-bit modified radix-2 Booth multipliers and supports reconfigurable operation across multiple precision modes, similiar to\cite{RPE-TCASII'24}. Exponent comparison and regime processing are implemented using a simplex multiplexer-based architecture. The datapath dynamically reallocates resources to balance throughput and precision, enabling efficient execution for varying operand widths. Var-FxP4 uses 4-bit multipliers, Var-FP8 combines 4-bit exponent and mantissa processing, while Posit8, Var-FxP8, and BF16 utilize 8-bit multipliers. Exponent outputs and mantissa products are realigned via a Quire accumulation. It also contains the simultaneous exponent comparison logic and partial products to fed to signed adder tree.

\textbf{Two-stage accumulation: }
Stage III, exponent difference from prior stage are aligened agaisnt the maximum exponent. Depending on sign bits, 2's complement is applied to the partial products. These aligned values are passed through a shifter and into a two-stage adder tree: Stage I (S3) implements a 4:2 Carry-Save Adder (CSA), while Stage II (S4) finalizes the summation using an additional CSA and a Carry-Select Adder (CSLA). Depending on the accumulation configuration, this stage ensures correct handling of partial sums and supports an optional Kulisch-style accumulation \cite{MACC-AMD}. This technique enables zero-passing based on the sign bit and selectively skips accumulation operations, thereby accelerating training.

\begin{figure}
    \centering
    \includegraphics[width=0.775\columnwidth,height=68mm]{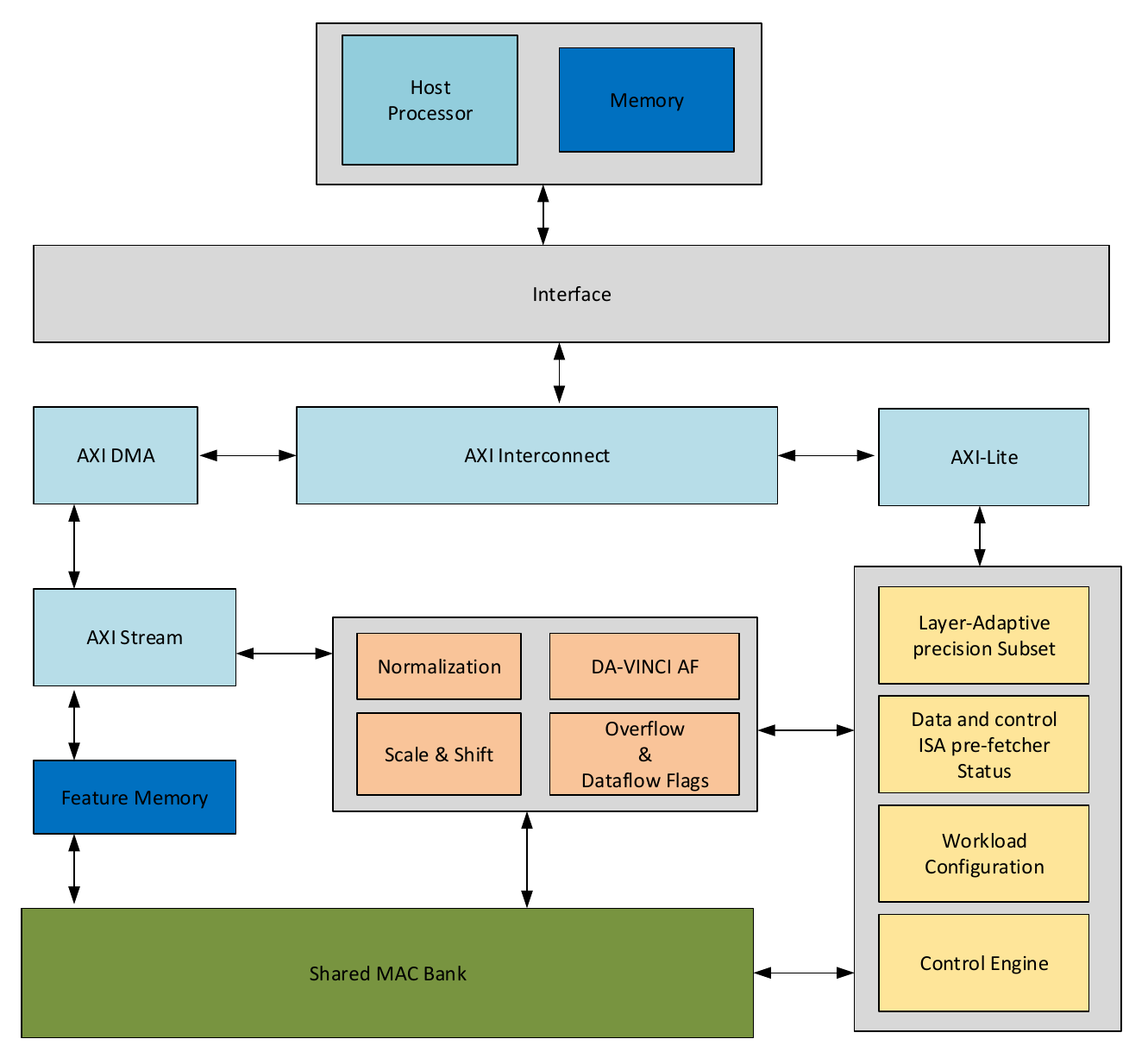}
    \vspace{-3mm}
    \caption{Proposed Precision-aware On-device Learning and Adaptive Runtime-cONfigurable (POLARON) edge-AI Engine.}
    \label{Fig-proposed-arch}
    \vspace{-5 mm}
\end{figure}

\textbf{Output Inter-format Restructuring:}
This stage performs normalization and rounding on the accumulated output. A Leading Zero Anticipator (LZA) determines the required shift for normalization, followed by adjustment of the exponent and mantissa. Rounding is carried out using the RoundTowardPositive method\cite{Mao-TVLSI’22}, which simplifies hardware implementation and mitigates underestimation errors caused by exponent-induced truncation. The final output is composed of sign, exponent, and mantissa fields in floating-point mode, or a regularized mantissa in fixed-point mode. 

\subsection{WILD-Q Lite Quantization Algorithm}
Prior works \cite{TSUNAMI, QuaRL, Flex-PE, Mega-mini, MSDF-MAC} have demonstrated on-device implementations of various AI workloads using precision-aware formats like DNN inference/training with Var-FxP4/8; Encoder/Decoder with Var-FxP8; RNNs, LSTMs, and Transformers with FP8; reinforcement learning with FxP8/16. Additionally, Posit-8 ($\sim$FP16), Posit-16 ($\sim$FP32), and BF16 have been shown to offer sufficient precision\cite{LPRE} for a wide range of AI workloads on hardware. The Hybrid Layer-Adaptive Precision Framework~\cite{Flex-PE} enables efficient AI execution across heterogeneous edge platforms by decoupling neural architecture search (NAS) from the training phase and supporting the deployment of specialized sub-networks through structured pruning of depth, width, kernel size, and input resolution - Awithout requiring retraining from scratch. Building on this principle, we propose a unified, precision-aware execution framework that supports a broad range of AI workloads, governed by a layer-adaptive precision subset. The proposed architecture employs a mixed-precision strategy integrated within the \textbf{PARV-CE} (Precision-Aware Runtime-Vectorized Compute Engine), which achieves competitive accuracy with less than 1.8\% degradation while significantly enhancing resource efficiency. Specifically, the framework utilizes Var-FxP4/8 for DNN inference and training, Var-FxP8 for encoder-decoder models, FP8 for RNNs, LSTMs, and Transformers, and FxP8/16 for reinforcement learning. In addition, it supports Posit-8 (comparable to FP16), Posit-16 (comparable to FP32), and BF16 for general-purpose AI tasks.

The quantization-aware training (QAT), incorporating layer-wise precision sensitivity analysis is adapted to preserve application accuracy. For the $l^\text{th}$ layer, the sensitivity metric $s_l$ to quantify the impact of precision on gradient magnitude:
\vspace{-4mm}

\begin{equation}
    s_{l_{\text{sc},k}} = \frac{\left( ||Q^{\text{Ad-P}}(\mathbf{w}_l) - \mathbf{w}_l|| - ||Q^{\text{Ad-P'}}_{\text{sc},k}(\mathbf{w}_l) - \mathbf{w}_l|| \right) \times ||\nabla \mathcal{L}_{\mathbf{w}_l}||}{n_l}
    \vspace{-2mm}
\end{equation}
with
\begin{equation}
    s_l = \max(s_{l_{\text{sc},8}}, s_{l_{\text{sc},4}}).
\end{equation}

\noindent where $Q^{\text{Ad-P}}(\cdot)$ denotes the adaptive precision quantization function, $\nabla \mathcal{L}_{\mathbf{w}_l}$ is the loss gradient with respect to weights $\mathbf{w}_l$, and $n_l$ is the number of weights in layer $l$. Unlike traditional symmetric ranges (e.g., $[-1,1]$), the scheme dynamically adjusts the saturation thresholds $W_l$ and $W_h$ based on the learned weight distribution. The quantization process involves computing a scale factor:
\vspace{-4mm}

\begin{equation}
\text{scale (k)} = \text{mean}(|W|) \cdot \frac{2^n - 1}{2^{n-1}},
\end{equation}
\begin{equation}
\widehat{W} = \text{round} \left( \left( \text{clip} \left( \frac{W}{k}, W_l, W_h \right) - W_l \right) \cdot \frac{2^n - 1}{W_h - W_l} \right),
\end{equation}
\begin{equation}
Q^{\text{Ad-P}}(W) = \widehat{W} \cdot \frac{W_h - W_l}{2^n - 1} + W_l,
\end{equation}

where $n$ denotes the quantization bit-width and $W$ is the original weight value. The resulting quantized representation is aligned with the model’s statistical characteristics, enabling efficient deployment across variable-precision hardware, similar to \cite{Flex-PE, LP-mixedge}. The framework leverages the parameterized clipping activation (PACT) function, which enables learnable thresholding during training:
\vspace{-4mm}

\begin{equation}
    \vspace{-2mm}
    y = \text{PACT}(x) = 0.5 (|x| - |x - \alpha| + \alpha).
    \vspace{-1mm}
\end{equation}

\begin{equation}
    x^q = \text{round} \left( y \times \frac{2^n - 1}{\alpha} \right) \times \frac{\alpha}{2^n - 1}.
\end{equation}

where $\alpha$ is the learnable clipping parameter. This approach mitigates quantization-induced accuracy loss by maintaining a balanced distribution around zero, while QAT compensates for approximation errors, preserving higher precision only in critical layers. Overall, the proposed framework enables efficient AI deployment on edge devices by combining quantization-aware training, entropy-based bit-width allocation, and layer-wise precision tuning to achieve high accuracy with improved resource efficiency.

\subsection{POLARON Architecture}
The proposed POLARON architecture presents a precision-adaptive accelerator designed to support efficient execution of AI workloads on edge platforms. At the heart of the design is a shared MAC bank, which performs varying precision MAC compute in Var-FxP4/8/16, Var-FP8, BF16, and Posit-8/16. Data ingress and egress are managed through an AXI interconnect, which bridges communication between the host processor/memory subsystem and the accelerator via AXI-Lite for configuration and AXI-DMA/Stream for high-throughput feature transfers. Incoming data is temporarily stored in the Feature Memory, to stage before fedding to MAC units. The architecture supports runtime adaptability via a layer-adaptive precision subset, enabling per-layer precision tuning for optimal trade-offs between accuracy and efficiency.

Post-processing with separate block includes normalization, scale-and-shift, overflow detection, and a custom CORDIC-driven DA-VINCI AF unit, which supports Swish, SoftMax, SeLU, GELU, Sigmoid, Tanh, ReLU. These modules ensure dynamic range alignment, prevent numerical saturation, and enable nonlinear transformations required for deep neural network layers. Control and coordination are managed by a runtime control engine that interfaces with a workload configuration module and a data/control ISA pre-fetcher, allowing preemptive adjustment of execution parameters based on layer-specific metadata. This tightly coupled data-flow-centric design ensures minimal latency, efficient memory utilization, and precision-aware computation tailored to the demands of modern AI workloads.

\section{Performance Analysis}
The performance evaluation of the proposed DNN accelerator follows a comprehensive co-design methodology, integrating software-based accuracy analysis with hardware-aware architectural modeling. An iso-functional emulation of the custom MAC arithmetic was implemented in Python 3.0 using the FxP-Math package and QKeras 2.3, executed within a Jupyter Notebook environment on an NVIDIA V100 GPU. The evaluation spans multiple precision and format with FP32 baseline. A diverse set of models was assessed under edge inference scenarios. The results demonstrate that the proposed WILD-QLite quantization framework sustains high model fidelity, with less than 1.8\% degradation in accuracy and a Quality of Results (QoR) score exceeding 98.5\%, validating its suitability for precision-constrained edge deployment.

The proposed PARV-CE and POLARON architecture were modeled in Verilog HDL with programmable support for multiple precision modes, demonstrating the flexibility. Functional verification was conducted using the QuestaSim simulator, with outputs cross-validated against the Python-based software emulation framework. FPGA synthesis and implementation were performed using the AMD Vivado Design Suite, and detailed resource utilization results are provided in Table~\ref{tab:mac-fpga} and Table~\ref{arch-comp}. Additionally, all designs were synthesized using Synopsys Design Compiler targeting the TSMC 28nm technology node, with corresponding post-synthesis performance metrics reported in Table~\ref{tab:mac-asic}, Table~\ref{stage-wise-comp} and Table~\ref{tab:arch-perf}. To ensure a fair comparison, several prior SoTA designs were re-implemented under identical experimental conditions. Results from both FPGA and ASIC implementations confirm that the proposed architecture achieves significant improvements in performance and resource efficiency over existing approaches.

\begin{table}[!t]
\caption{FPGA Resource utilization for SoTA MAC designs}
\label{tab:mac-fpga}
\renewcommand{\arraystretch}{1.25}
\resizebox{\columnwidth}{!}{%
\begin{tabular}{c|c|ccccc}
\hline
\multirow{2}{*}{Design} & \multirow{2}{*}{Precision} & \multicolumn{5}{c}{FPGA Utilization (Virtex 707)} \\ \cline{3-7} 
 &  & \multicolumn{1}{c|}{LUTs} & \multicolumn{1}{c|}{FFs} & \multicolumn{1}{c|}{Delay (us)} & \multicolumn{1}{c|}{Power (mW)} & \begin{tabular}[c]{@{}c@{}}Arith. Intensity \\ (pJ/Op)\end{tabular} \\ \hline
\multirow{2}{*}{AMD IP (FPGA)} & FxP4 & \multicolumn{1}{c|}{53} & \multicolumn{1}{c|}{28} & \multicolumn{1}{c|}{3.09} & \multicolumn{1}{c|}{3.48} & 10.75 \\ \cline{2-7} 
 & FxP8 & \multicolumn{1}{c|}{130} & \multicolumn{1}{c|}{44} & \multicolumn{1}{c|}{3.816} & \multicolumn{1}{c|}{7.26} & 27.7 \\ \cline{2-7} 
\cite{GR-TRETS'23} & FxP16 & \multicolumn{1}{c|}{369} & \multicolumn{1}{c|}{76} & \multicolumn{1}{c|}{9.051} & \multicolumn{1}{c|}{16.9} & 153 \\ \cline{2-7} 
 & FxP32 & \multicolumn{1}{c|}{1426} & \multicolumn{1}{c|}{214} & \multicolumn{1}{c|}{5.931} & \multicolumn{1}{c|}{22} & 130.4 \\ \hline
\multirow{4}{*}{TVLSI'25\cite{Flex-PE}} & FxP8 & \multicolumn{1}{c|}{256} & \multicolumn{1}{c|}{224} & \multicolumn{1}{c|}{5.98} & \multicolumn{1}{c|}{9.23} & 55.2 \\ \cline{2-7} 
 & FxP16 & \multicolumn{1}{c|}{427} & \multicolumn{1}{c|}{369} & \multicolumn{1}{c|}{6.5} & \multicolumn{1}{c|}{11.78} & 76.57 \\ \cline{2-7} 
 & FxP32 & \multicolumn{1}{c|}{681} & \multicolumn{1}{c|}{745} & \multicolumn{1}{c|}{7.34} & \multicolumn{1}{c|}{31} & 227.54 \\ \cline{2-7} 
 & SIMD-Pipelined & \multicolumn{1}{c|}{897} & \multicolumn{1}{c|}{1231} & \multicolumn{1}{c|}{11.7} & \multicolumn{1}{c|}{59.4} & 694 \\ \hline
\multirow{4}{*}{ISCAS'25\cite{LPRE}} & Posit8 & \multicolumn{1}{c|}{467} & \multicolumn{1}{c|}{175} & \multicolumn{1}{c|}{2.68} & \multicolumn{1}{c|}{68} & 182.24 \\ \cline{2-7} 
 & Posit16 & \multicolumn{1}{c|}{2083} & \multicolumn{1}{c|}{528} & \multicolumn{1}{c|}{4.35} & \multicolumn{1}{c|}{189} & 822.15 \\ \cline{2-7} 
 & Posit32 & \multicolumn{1}{c|}{6813} & \multicolumn{1}{c|}{806} & \multicolumn{1}{c|}{8} & \multicolumn{1}{c|}{347} & 2776 \\ \cline{2-7} 
 & SIMD-L.Posit & \multicolumn{1}{c|}{4613} & \multicolumn{1}{c|}{2078} & \multicolumn{1}{c|}{6.2} & \multicolumn{1}{c|}{276} & 426 \\ \hline
TCAS-II'24\cite{RPE-TCASII'24} & SIMD-INT4/FP8/16/32 & \multicolumn{1}{c|}{8054} & \multicolumn{1}{c|}{1718} & \multicolumn{1}{c|}{4.62} & \multicolumn{1}{c|}{296} & 152 \\ \hline
TVLSI'23\cite{Mao-TVLSI’22} & SIMD-FP16/32/64 & \multicolumn{1}{c|}{8065} & \multicolumn{1}{c|}{1072} & \multicolumn{1}{c|}{5.56} & \multicolumn{1}{c|}{378} & 543 \\ \hline

\multirow{4}{*}{Access'24\cite{QuantMAC}} 
& Q-4b & \multicolumn{1}{c|}{24} & \multicolumn{1}{c|}{16} & \multicolumn{1}{c|}{0.98} & \multicolumn{1}{c|}{2.2} & 2.16 \\ \cline{2-7} 

& Q-8b & \multicolumn{1}{c|}{52} & \multicolumn{1}{c|}{88} & \multicolumn{1}{c|}{1.57} & \multicolumn{1}{c|}{6.36} & 10 \\ \cline{2-7} 

 & Q-16b & \multicolumn{1}{c|}{106} & \multicolumn{1}{c|}{168} & \multicolumn{1}{c|}{2.2} & \multicolumn{1}{c|}{11.77} & 26 \\ \cline{2-7} 
 &  \multicolumn{1}{c|}{\begin{tabular}[c]{@{}c@{}}Vector Q-MAC\\ (FxP-8/16/32)\end{tabular}} & \multicolumn{1}{c|}{1502} & \multicolumn{1}{c|}{2418} & \multicolumn{1}{c|}{40.32} & \multicolumn{1}{c|}{21.38} & 108 \\ \hline

TCAS-I'22\cite{ILM} & FxP8 & \multicolumn{1}{c|}{238} & \multicolumn{1}{c|}{32} & \multicolumn{1}{c|}{2.75} & \multicolumn{1}{c|}{2.8} & 7.6 \\ \hline
\multirow{3}{*}{\begin{tabular}[c]{@{}c@{}}TRETS'23\\ CORDIC\cite{GR-TRETS'23}\end{tabular}} & FxP4 & \multicolumn{1}{c|}{35} & \multicolumn{1}{c|}{58} & \multicolumn{1}{c|}{1.406} & \multicolumn{1}{c|}{4.36} & 6.13 \\ \cline{2-7} 
 & FxP8 & \multicolumn{1}{c|}{54} & \multicolumn{1}{c|}{88} & \multicolumn{1}{c|}{1.518} & \multicolumn{1}{c|}{6.6} & 10 \\ \cline{2-7} 
 & FxP16 & \multicolumn{1}{c|}{95} & \multicolumn{1}{c|}{162} & \multicolumn{1}{c|}{2.124} & \multicolumn{1}{c|}{12.21} & 25.9 \\ \hline
Baseline & \multirow{2}{*}{\begin{tabular}[c]{@{}c@{}}Ad-FxP-4/8/16,\\ Var-FP8/16, Posit8/16\end{tabular}} & \multicolumn{1}{c|}{875} & \multicolumn{1}{c|}{1134} & \multicolumn{1}{c|}{10.89} & \multicolumn{1}{c|}{54.7} & 37.23 \\ \cline{1-1} \cline{3-7} 
Proposed &  & \multicolumn{1}{c|}{326} & \multicolumn{1}{c|}{182} & \multicolumn{1}{c|}{2.89} & \multicolumn{1}{c|}{31.67} & 5.72 \\ \hline
\end{tabular}}
\vspace{-3mm}
\end{table}

\begin{table}[!t]
\caption{ASIC Performance Comparison of SoTA MAC designs}
\renewcommand{\arraystretch}{1.1}
\centering
\resizebox{0.85\columnwidth}{!}{%
\label{tab:mac-asic}
\begin{tabular}{c|c|c|c|c|c}
\hline
\multirow{2}{*}{Design} & Tech. & Voltage & Freq & Area & Power \\ \cline{2-6} 
 & nm & V & GHz & mm\textsuperscript{2} & mW \\ \hline

TVLSI'25\cite{Flex-PE} & 28 & 0.9 & 1.36 & 0.049 & 7.3 \\\hline
ISCAS'25\cite{LPRE} & 28 & 0.9 & 1.12 & 0.024 & 32.68 \\\hline
TCAS-II'24\cite{RPE-TCASII'24} & 28 & 1 & 1.47 & 0.010 & 15.87 \\ \hline
TCAD'24\cite{Tan-TCAD'24} & 28 & 1 & 1.47 & 0.024 & 82.4 \\ \hline
TCAS-II'24\cite{Tan-TCASII'24} & 28 & 1 & 1.56 & 0.022 & 72.3 \\ \hline
TCAS-II'22\cite{Posit-FP-VMAC-TCASII'22} & 28 & 1.05 & 0.67 & 0.052 & 99 \\ \hline
TVLSI'23\cite{Tan-TVLSI'23} & 28 & 1 & 2.22 & 0.013 & 59.3 \\ \hline
TVLSI'22\cite{Mao-TVLSI’22} & 28 & 1 & 1.43 & 0.013 & 29.3 \\ \hline
Baseline & 28 & 0.9 & 1 & 0.062 & 112 \\ \hline
\textbf{Proposed} & 28 & 0.9 & 1.86 & 0.011 & 28.2 \\ \hline
 \end{tabular}}
 \vspace{-3mm}
\end{table}

\begin{table*}[!t]
 \caption{comparison for Stage-wise resources consumed by different MAC Units}
\vspace{-2mm}
\label{stage-wise-comp}
\renewcommand{\arraystretch}{1.15}
\resizebox{\textwidth}{!}{%
\begin{tabular}{c|cc|cc|cc|cc|cc|cc}
\hline
 & \multicolumn{2}{c|}{TVLSI'22\cite{Mao-TVLSI’22}} & \multicolumn{2}{c|}{TCAS-II'22\cite{Posit-FP-VMAC-TCASII'22}} & \multicolumn{2}{c|}{TVLSI'23\cite{Tan-TVLSI'23}} & \multicolumn{2}{c|}{TCAS-II'24\cite{Tan-TCASII'24}} & \multicolumn{2}{c|}{TCAD'24\cite{Tan-TCAD'24}} & \multicolumn{2}{c}{Proposed} \\ \hline
Stage & \multicolumn{1}{c|}{Area ($\mu$m\textsuperscript{2})} & Power(mW) & \multicolumn{1}{c|}{Area ($\mu$m\textsuperscript{2})} & Power(mW) & \multicolumn{1}{c|}{Area ($\mu$m\textsuperscript{2})} & Power(mW) & \multicolumn{1}{c|}{Area ($\mu$m\textsuperscript{2})} & Power(mW) & \multicolumn{1}{c|}{Area ($\mu$m\textsuperscript{2})} & Power(mW) & \multicolumn{1}{c|}{Area ($\mu$m\textsuperscript{2})} & Power(mW) \\ \hline

Input Pre-proc. & \multicolumn{1}{c|}{1902} & 14.5 & \multicolumn{1}{c|}{8079} & 16.2 & \multicolumn{1}{c|}{\multirow{2}{*}{6575}} & \multirow{2}{*}{24.5} & \multicolumn{1}{c|}{\multirow{2}{*}{13432}} & \multirow{2}{*}{41} & \multicolumn{1}{c|}{\multirow{2}{*}{14735}} & \multirow{2}{*}{45} & \multicolumn{1}{c|}{673} & 1.78 \\ \cline{1-5} \cline{12-13}

\begin{tabular}[c]{@{}c@{}}Mantissa Mult.\& \\ Exponent Proc.\end{tabular} & \multicolumn{1}{c|}{5907} & 45 & \multicolumn{1}{c|}{22772} & 43.5 & \multicolumn{1}{c|}{} & & \multicolumn{1}{c|}{} & & \multicolumn{1}{c|}{} & & \multicolumn{1}{c|}{3978} & 9.67 \\ \hline

Accumulation & \multicolumn{1}{c|}{2810} & 22 & \multicolumn{1}{c|}{13273} & 26 & \multicolumn{1}{c|}{1540} & 8.7 & \multicolumn{1}{c|}{5636} & 20 & \multicolumn{1}{c|}{3058} & 12 & \multicolumn{1}{c|}{3426} & 9.2 \\ \hline

Output Proc. & \multicolumn{1}{c|}{2778} & 21.2 & \multicolumn{1}{c|}{5855} & 11 & \multicolumn{1}{c|}{4914} & 26 & \multicolumn{1}{c|}{2849} & 11.4 & \multicolumn{1}{c|}{6320} & 25.5 & \multicolumn{1}{c|}{3068} & 7.52 \\ \hline

Total & \multicolumn{1}{c|}{13397} & 102.7 & \multicolumn{1}{c|}{49979} & 96.7 & \multicolumn{1}{c|}{13029} & 59.2 & \multicolumn{1}{c|}{21917} & 72.4 & \multicolumn{1}{c|}{24113} & 82.5  & \multicolumn{1}{c|}{11147} & 28.2 \\ \hline

Op. Freq. (GHz)& \multicolumn{2}{c|}{1.42} & \multicolumn{2}{c|}{0.67} &\multicolumn{2}{c|}{2.22} &\multicolumn{2}{c|}{1.56} &\multicolumn{2}{c|}{1.47} &\multicolumn{2}{c}{1.86} \\\hline

\end{tabular}
}
\vspace{-3mm}
\end{table*}

\begin{table*}[!t]
\caption{FPGA resources comparison with SoTA accelerator architectures\cite{XXie-TCASI'22, DThanh-TCASI'22, GR-TRETS'23, VaPr-WLee-TVLSI'23, BWu-TCAS-I'24, Flex-PE}}

\vspace{-2mm}
\label{arch-comp}
\renewcommand{\arraystretch}{1.05}
\resizebox{\linewidth}{!}{%
\begin{tabular}{c|c|c|c|c|c|c|c|c|c|c}
\hline
 & TCAS-I'22\cite{XXie-TCASI'22} & TCAS-I'22\cite{DThanh-TCASI'22} & TCAD'23\cite{WJiang-TCAD'23} & TRETS'23\cite{GR-TRETS'23} & TCAS-II'23\cite{SKi-TCASII'23} & TVLSI'23\cite{VaPr-WLee-TVLSI'23}  & TCAS-I'24\cite{BWu-TCAS-I'24} & TCAS-I'24\cite{MKim-TCAS-I'24} & TVLSI'25\cite{Flex-PE} & Proposed \\ \hline
 
Platform  & Intel Aria-10 & KCU15 & ZCU-102 & VC707 & XCVU9P & ZCU102 & ZU3EG & A7-100T & VC707 & VC707\\ \hline
Model  & MobileNet-v2 & YoloV3-tiny & MobileNet-v2 & LeNet-5 & YoloV3-tiny & XoR-Net & ResNet-50 & YoloV3-tiny & VGG-16 & YoloV3-tiny \\ \hline
Precision  & 8 & 8 & 8 & 8/16 & 8 & 8 & 8 & 8 & 8/16/32 & 4/8/16 \\ \hline
LUTs  & 102.6 K & 213.3 K & 164.4. K & 144 K & 132 K & 117 K & 40.78 K & 50.2 K & 38.7 K & \textbf{37.2} K \\ \hline
FFs  & - & 352 K & - & 155.8 K & 39.5 K & 74 K & 45.25 K & 58.1 K & 7.4 K & \textbf{8.6} K \\ \hline
DSPs & 512 & 2240 & 1283 & 23 & 96 & 132  & 257 & 240 & 73 & - \\ \hline
Freq (MHz)  & 170 & 200 & 333 & 466 & 150 & 300 & 150 & 100 & 466 & 250 \\ \hline
Power (W) & 4.6 & 0.51 & 0.96 & 1.54 & 5.52 & 6.58 & 1.4 & 2.2 & 2.24 & 0.93 \\ \hline
\end{tabular}}
\vspace{-3mm}
\end{table*}

The proposed PARV-CE shows significant improvements in efficiency and scalability across both FPGA and ASIC platforms. On FPGA (Table \ref{tab:mac-fpga}), demonstrates lower resource utilization and power consumption while delivering competitive throughput, outperforming SoTA designs with almost 2$\times$ in PDP and 1.5$\times$ in arithmetic intensity. Compared to SIMD-based and high-precision MAC implementations, the proposed design offers broader precision support and maintains consistently high efficiency with minimal logic overhead. On ASIC (Table \ref{tab:mac-asic}), PARV-CE achieved over 1.3$\times$ quicker operating frequency and consume 2$\times$ smaller area compared to SoTA works, with total consumption reduced by 40–50\% relative to prior precision-flexible MACs. Stage-wise analysis (Table \ref{stage-wise-comp}) further emphasizes these advantages, predominantly arising from a reduction in both area and power across major pipeline stages. This effectively highlights the proposed PARV-CE for compact, scalable and faster edge execution.

The effective throughput benefits from PARV-CE and adaptive precision frameworks translate very significantly to POLARON accelerator implementation. As shown in Table \ref{arch-comp}, the FPGA-based design consumes 3$\times$ lesser LUTs and FFs compared to SoTA designs, at similiar frequency. The use of a trans-precision compute engine enables broad support for mixed-precision formats, including FxP4/8/16 and Posit8/16, which helps reduce logic complexity and improves runtime adaptability. In terms of system-level evaluation (Table \ref{tab:arch-perf}), POLARON shows 2-4$\times$ better energy efficiency compared to SoTA designs and better accuracy. The ODT for VGG-16 was observed 4.2\% times faster compared to prior works\cite{Flex-PE}. Overall, the proposed architecture offers a compelling blend of flexibility, accuracy, and efficiency making it an ideal candidate for low-power edge-AI deployments.

The proposed POLARON was deployed for object detection and classification tasks on the Pynq-Z2 platform, utilizing the ARM Cortex-A9 processor as the host CPU. Real-time deployment demonstrates that the proposed system outperforms prior works in both latency and power consumption, achieving 11 fps and 0.8W W on Pynq-Z2, compared to 186.4 ms / 2.24 W for \cite{Flex-PE} on VC707, 772 ms / 1.524 W for \cite{GR-TRETS'23} on VC707, 184 ms / 0.93 W for \cite{LPRE} on Pynq-Z2, and baseline values of 226 ms / 1.34 W for Jetson Nano and 555 ms / 2.7 W for Raspberry Pi. The architecture also benefits from early-exit strategy that conditionally skips layers. The inference accuracy for RNNs, LSTMs stays within 1.2\% compared to FP32 and RL between 1\%. We mark this as a opportunity for future exploration. 

\begin{table*}[!t]
\caption{Evaluation Metrics Comparison between diverse SoTA AI accelerator Designs}
\label{tab:arch-perf}
\renewcommand{\arraystretch}{1.05}
\resizebox{\textwidth}{!}{%
\begin{tabular}{c|c|c|c|c|c|c|c|c}
\hline
Design & Network Topology & Data-type & Tech & Freq (MHz) & Area (mm\textasciicircum{}2) & Power (W) & \begin{tabular}[c]{@{}c@{}}Energy Efficiency\\ (TOPS/W)\end{tabular} & Accuracy \\ \hline
\multirow{2}{*}{JSSC'25\cite{VSA-JSSC'25}} & Vector Systolic Array & \multirow{2}{*}{FxP4/8} & \multirow{2}{*}{28nm} & 172 & 1.04 & 0.6 & 8.33 & 71.68 \\ \cline{2-2} \cline{5-9} 
 & G-VSA &  &  & 199 & 2 & 0.3 & 3.26 & 67.2 \\ \hline
 
\multirow{2}{*}{TVLSI'25\cite{Flex-PE}} & VGG-16 & \multirow{2}{*}{FxP4/8/16/32} & VC707 (28nm) & 466 & - & 2.24 & 8.42 & 84.6 \\ \cline{2-2} \cline{4-9} 
 & Systolic Array (8x8) &  & TSMC-28nm & 1435 & 1.8 & 0.53 & 10.83 & - \\ \hline
\multirow{2}{*}{TVLSI'25\cite{MSDF-MAC}} & 784-200-100-10 & \multirow{2}{*}{FxP8} & \multirow{2}{*}{ASIC-45nm} & \multirow{2}{*}{588} & 6.13 & 0.6 & 1.48 & 97.4 \\ \cline{2-2} \cline{6-9} 
 & 784-256-10 &  &  &  & 5.86 & 0.642 & 1.39 & 96.73 \\ \hline
\multirow{2}{*}{ISCAS'25\cite{LPRE}} & LeNet-5 & \multirow{2}{*}{Posit-8/16/32} & VCU129 (16nm) & 46.35 & - & 0.93 & 4.54 & 98.7 \\ \cline{4-9} 
 & (64 PQRE) &  & ASU-7nm & 1835 & 0.63 & 1.2 & 6.2 & - \\ \cline{4-9}
 &  &  & TSMC-28nm & 1350 & 2.32 & 2.8 & 1.98 & - \\ \hline
\multirow{2}{*}{Access'24 \cite{QuantMAC}} & LeNet-5 & \multirow{2}{*}{FxP8} & Virtex-7 & 262 & - & 0.18 & 3.78 & 97.2 \\ \cline{4-9} 
 & (64 Quant-MAC) &  & CMOS 28nm & 1382 & 1.4 & 1.86 & 0.78 & - \\ \hline
TCAD'23 & Tiny-Yolo v3 & FxP8 & XCVU9P & 150 & - & 5.52 & - & 72.98 \\ \hline
ISSCC'23\cite{DR-ISSCC'23} & ResNet-20 & FP16/32, BF16 & 22nm FDX & 420 & 1.9 & 0.123 & 1.66 & 92.2 \\ \hline
ISCAS'24\cite{ODL-ISCAS'24} & ResNet-50 & FxP4/FP-16/32 & 28nm & 160 & 1.84 & 67.4 & 2.19 & 77.56 \\ \hline
TVLSI'23 & XNOR-Net & 8b & ZCU102 & 300 & - & 6.58 & 1.52 & 62.98 \\ \hline
\begin{tabular}[c]{@{}l@{}}TCAS-I'22\cite{PL-NPU_TCASI'22} \\ (PL-NPU)\end{tabular} & ResNet-18 & Posit8 & 28nm & 1040 & 5.28 & 343 & 1.625 & 70.1  \\ \hline
\multirow{2}{*}{TRETS'23\cite{GR-TRETS'23}} & \multirow{2}{*}{196-64-32-32-10} & \multirow{2}{*}{FxP8} & VC707 (28nm) & 466 & - & 1.524 & 4.5 & 95.06 \\ \cline{4-9} 
 &  &  & CMOS-28nm & 826 & 2.34 & 0.89 & 9.28 & - \\ \hline
 \multirow{2}{*}{TCAS-I'22\cite{ILM}} & \multirow{2}{*}{196-64-32-32-10} & \multirow{2}{*}{FxP8} & VC707 (28nm) & 357 & - & 1.78 & 8.02 & 96.59 \\ \cline{4-9} 
 &  &  & CMOS-28nm & 1376 & 1.04 & 0.62 & 12.22 & - \\ \hline
 \multirow{4}{*}{Baseline} & \multirow{2}{*}{LeNet-5 (64 PARV-CE)} & \multirow{2}{*}{Ad-FxP-4/8/16,} & VC707 (28nm) & 196 & - & 1.38 & 2.27 & 98.6 \\ \cline{4-9} 
 &  &  & TSMC-28nm & 968 & 1.23 & 1.68 & 8.2 & - \\ \cline{2-2} \cline{4-9} 
 & \multirow{2}{*}{VGG-16 (256 PARV-CE)} & \multirow{2}{*}{Var-FP8/16, Posit8/16} & VC707 (28nm) & 204 & - & 1.46 & 2.23 & 68.7 \\ \cline{4-9} 
 &  &  & TSMC-28nm & 1167 & 3.85 & 1.8 & 12.8 & - \\ \hline

\multirow{4}{*}{Proposed} & \multirow{2}{*}{LeNet-5 (64 PARV-CE)} & \multirow{2}{*}{Ad-FxP-4/8/16,} & VC707 (28nm) & 250 & - & 0.93 & 7.57 & 98.3 \\ \cline{4-9} 
 &  &  & TSMC-28nm & 1023 & 0.64 & 0.88 & 15.2 & - \\ \cline{2-2} \cline{4-9} 
 & \multirow{2}{*}{VGG-16 (256 PARV-CE)} & \multirow{2}{*}{Var-FP8/16, Posit8/16} & VC707 (28nm) & 237 & - & 1.16 & 5.86 & 68.7 \\ \cline{4-9} 
 &  &  & TSMC-28nm & 1238 & 2.2 & 1.2 & 16.5 & - \\ \hline
\end{tabular}}
\vspace{-5mm}
\end{table*}

\section{Conclusion}

This work presents POLARON, a precision-aware and runtime-adaptive edge AI accelerator, incorporating the PARV-CE to address the emerging needs of energy-efficient, scalable, and trans-precision AI computation at the edge. It supports unified MAC operations across Var-FxP4/8/16, Var-FP8/16, BF16, and Posit8/16, for diverse AI workloads. With the integration of the WILD-QLite quantization algorithm and layer-adaptive execution framework, the design achieves improved throughput and resource efficiency, while maintaining model accuracy within 1.8\%, compared to FP32 baseline. The comprehensive evaluation across FPGA and ASIC platforms demonstrates up to 2× improvement in PDP, 3× reduction in resource usage, and 4× better energy efficiency compared to SoTA designs. The real-time deployment on the Pynq-Z2 platform confirms the suitability of POLARON for low-power edge inference and training. The empirical assessment place POLARON in next-generation AI systems, to exploit precision scalability, hardware adaptability, and performance efficiency under strict edge constraints.

\bibliographystyle{ieeetr}
\bibliography{conference_101719}

\end{document}